# Efficient optical pumping of alkaline atoms for evanescent fields at dielectric-vapor interfaces

*Eliran Talker[1], Pankaj Arora[1], Yefim Barash[1], David Wilkowski[2,3,4], and Uriel Levy[1,*]*
[1]*Department of Applied Physics, The faculty of Science, The Center for Nanoscience and Nanotechnology, The Hebrew University of Jerusalem, Jerusalem, 91904, Israel*
[2]*School of Physical and Mathematical Sciences, Nanyang Technological University, 637371 Singapore*
[3]*Centre for Quantum Technologies, National University of Singapore, 117543 Singapore*
[4]*MajuLab, CNRS-UCA-SU-NUS-NTU International Joint Research Unit, Singapore*
*\*Email id: ulevy@mail.huji.ac.il*

## Abstract:

We experimentally demonstrate hyperfine optical pumping of rubidium atoms probed by an evanescent electromagnetic field at a dielectric-vapor interface. This light-atom interaction at the nanoscale is investigated using a right angle prism integrated with a vapor cell and excited by evanescent wave under total internal reflection. An efficient hyperfine optical pumping, leading to a complete suppression of absorption on the probed evanescent signal, is observed when a pump laser beam is sent at normal incidence to the interface. In contrast, when the pump and probe beams are co-propagating in the integrated prism-vapor cell, no clear evidence of optical pumping is observed. The experimental results are supported by a detailed model based on optical Bloch equation of a four atomic levels structure. The obtained on-chip highly efficient optical pumping at the nanoscale is regarded as an important step in the quest for applications such as optical switching, magnetometry and quantum memory.

## Introduction:

The reflected signal from a dielectric-vapor interface exhibits resonant behavior around the resonance line of the atoms [1]. As light is mainly reflected from a thin atomic layer of subwavelength thickness, this method is sensitive to atom-surface interactions [2,3]. Moreover, only atoms flying within this layer will contribute to the reflected signal, whereas atoms with velocity component normal to the interface rapidly leave the interacting zone. This peculiarity leads to a selective reflection effect both for total internal reflection and normal incidence schemes [4–9]. Light-atoms interactions at the nanoscale with evanescent wave has led to numerous applications such as atomic mirror, magnetometer, optical modulation, optical switching and other nonlinear effects were demonstrated [9–19].

Achieving high contrast hyperfine optical pumping with hot vapors is a challenging task due to the Doppler broadening effect. If the pump and probe beams are co/counter propagating, optical pumping is achieved only for a pre-selected atomic velocity class in resonance with the pump laser beam [20]. Thus, other atomic velocity classes cannot be pumped. If, on the other hand, a cross beam configuration is used, atoms can be pumped regardless of their velocity component along the probe beam. Still, atoms with a velocity component along the pump propagation direction (perpendicular to the probe), which do not match the pump detuning, will not be pumped. In this work, we take advantage of velocity selection process at the nanoscale to remove this obstacle and demonstrate both theoretically and experimentally efficient hyperfine optical pumping on the reflected evanescent field signal within a cross beam pump-probe scheme. A right angle prism integrated with a vapor cell has been used to support evanescent wave under total internal reflection scheme in the vapor [21]. Optical pumping is observed in the selective reflection signal when the propagating pump and evanescent probe beams are orthogonal to each other. Similar experimental configurations have been used to explore atom-surface scattering properties [22,23], reflection spectroscopy of a spin-polarized gas [24] and direct observation of the evanescent field [25]. To reveal the optical pumping mechanism in the subwavelength atomic layer, a detailed model based on the optical Bloch equation is developed and compared to our experimental results.

## Numerical Results:

Figure 1 (a) shows the fine and hyperfine energy level scheme of $^{85}$Rb used in the experiment. The $5^2S_{1/2}$ ground state is split into a hyperfine structure with a frequency spacing of $\Delta_{HFS} = 3.035$ GHz. The pump laser is tuned on the $F_g$ = 2, $5^2S_{1/2} \rightarrow F_e$ = 3, $5^2P_{1/2}$ D$_1$ transition at 795 nm, whereas the probe laser is scanned over the all $5^2S_{1/2} \rightarrow 5^2P_{3/2}$ D$_2$ transition at 780 nm.

The Doppler frequency broadening in the experiment allows only to resolve the ground state hyperfine level and the excited state fine level. For this reason, the complex level structure of the $^{85}$Rb atom is reduced to a four levels scheme as shown in Figure 1(b). This simplified level scheme is sufficient to investigate the optical pumping between the two hyperfine ground states, noted $|1\rangle, |2\rangle$. We keep the same energy splitting as for the $^{85}$Rb atom. Levels $|3\rangle$ and $|4\rangle$ are the excited state levels corresponding to D$_1$ transition and D$_2$ transition respectively. In the theoretical model, the pump beam is set on the $|1\rangle \rightarrow |3\rangle$ transition and the probe beam is set between $|1\rangle \rightarrow |4\rangle$ and $|2\rangle \rightarrow |4\rangle$ transitions, respectively. $\Delta_{pr}, \Delta_{pu}$ are the detunings and $\Omega_{pr}, \Omega_{pu}$ are the Rabi frequencies of the probe and pump beam respectively. The model also corresponds to our experimental configuration, shown later in Figure 2.

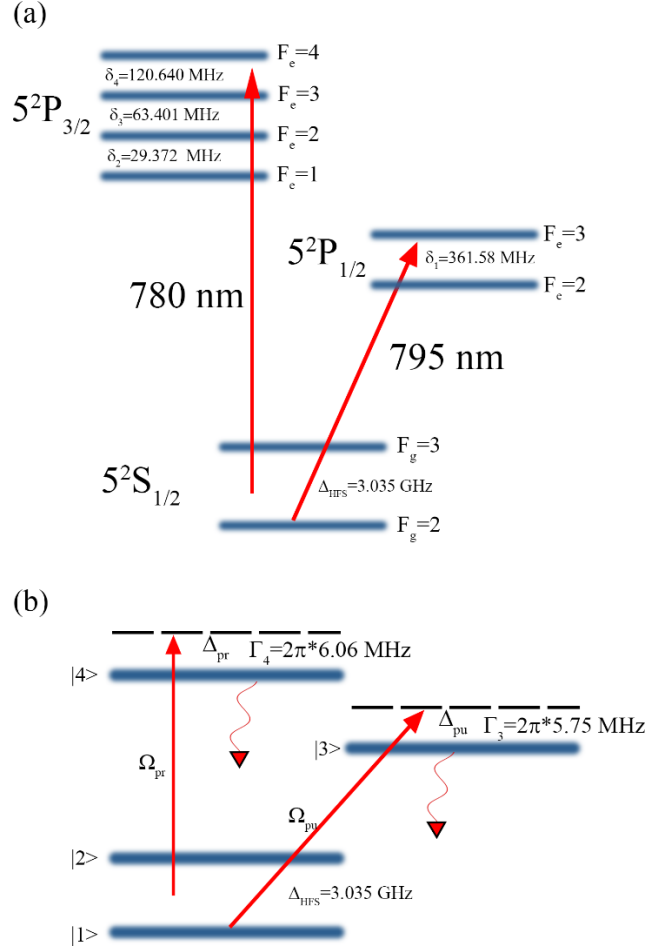

**Figure 1**: (a) Fine and hyperfine energy structure of $^{85}$Rb considered in the experiment (b) Simplified energy level scheme used for the theoretical model

To study the interaction of the electromagnetic field with atoms, we use the electric dipole interaction operator,

$$V = -\vec{d}_1 \cdot \vec{E}_{pu} - \vec{d}_2 \cdot \vec{E}_{pr} = -\vec{d}_1 \cdot \hat{\varepsilon}_{pu} E_{pu} \cdot e^{i(\omega_{pu}\cdot t + k_{pu}\cdot z)} - \vec{d}_2 \cdot \hat{\varepsilon}_{pr} E_{pr}(z) e^{i(\omega_{pr}\cdot t - k_{pr}\cdot x)},$$

where $\vec{d}_1$, and $\vec{d}_2$ are the dipole matrix element of the D$_1$, and D$_2$ transition respectively. $E_{pu}$, $\omega_{pu}$, $k_{pu}$ and $E_{pr}$, $\omega_{pr}$, $k_{pr}$ are the amplitude, frequency, and wavenumber of the pump and probe beams, respectively. $\hat{\varepsilon}_{pu} = \hat{z}$ and $\hat{\varepsilon}_{pr} = \hat{x}$ are the electric field polarizations of the pump, and probe beams, respectively. The evanescent nature of the probe beam implies that

$$E_{pr}(z) = E_{pr} \cdot \exp(-\kappa_{pr} z)$$

Where $\kappa_{pr} = \delta_{pr}^{-1}$, $\delta_{pr}$ is the penetration depth, is equal to:

$$\kappa_{pr} = \frac{\omega_{pr}}{c}\sqrt{n_1^2 \sin^2(\theta_i) - 1}$$

Here, $n_1$ is the refractive index of the prism and $\theta_i$ is the incidence angle of the probe laser beam, which is slightly larger than the critical angle for total internal reflection.

With these definitions in mind, the interaction operator $V$ in matrix form under rotating wave approximation can be written as:

$$V = -\hbar \begin{pmatrix} 0 & 0 & 0 & \Omega_{pr}(z) \\ 0 & 0 & \Omega_{pu} & \Omega_{pr}(z) \\ 0 & \Omega_{pu} & 0 & 0 \\ \Omega_{pr}(z) & \Omega_{pr}(z) & 0 & 0 \end{pmatrix}$$

Where

$$\Omega_{pr}(z) = -\frac{\vec{d}_2 \cdot \hat{\varepsilon}_{pr}}{\hbar} E_{pr}(z)$$

$$\Omega_{pu} = -\frac{\vec{d}_1 \cdot \hat{\varepsilon}_{pu}}{\hbar} E_{pu}$$

are the Rabi frequencies of the transitions of interest. The evolution of the atomic density operator is governed by the following time dependent optical Bloch equation:

$$\frac{d\rho}{dt} = -\frac{i}{\hbar}[\mathcal{H}_{tot}, \rho] + \mathcal{L}\rho$$

Here, the Hamiltonian $\mathcal{H}_{tot}$ is the sum of the interaction operator $V$ and the Hamiltonian of the free atom,

$$\mathcal{H}_0 = \begin{pmatrix} -\Delta_{HFS} & 0 & 0 & 0 \\ 0 & 0 & 0 & 0 \\ 0 & 0 & \Delta_{pu} & 0 \\ 0 & 0 & 0 & \Delta_{pr} \end{pmatrix}$$

$\mathcal{L}$ is the Lindblad relaxation operator,

$$\mathcal{L} = \frac{1}{2}\begin{pmatrix} 2(\Gamma_3\rho_{33} + \Gamma_4\rho_{44}) & 0 & -\Gamma_3\rho_{13} & -\Gamma_4\rho_{14} \\ 0 & 2(\Gamma_3\rho_{33} + \Gamma_4\rho_{44}) & -\Gamma_3\rho_{23} & -\Gamma_4\rho_{24} \\ -\Gamma_3\rho_{31} & -\Gamma_3\rho_{32} & -2\Gamma_3\rho_{33} & -\frac{1}{2}(\Gamma_3 + \Gamma_4)\rho_{34} \\ -\Gamma_4\rho_{41} & -\Gamma_4\rho_{42} & -\frac{1}{2}(\Gamma_3 + \Gamma_4)\rho_{43} & -2\Gamma_4\rho_{44} \end{pmatrix}$$

Here, $\Gamma_3$ and $\Gamma_4$ are the natural linewidth for levels $|3\rangle$, and $|4\rangle$ respectively as shown in Figure 1(b).

The steady-state regime of the optical Bloch equations is obtained keeping the spatial derivative term of the convective derivative $d/dt = \partial/\partial t + \vec{v}.\vec{\nabla}$. Since the system is supposed to have translational invariance symmetry in the *x-y* plane, the optical Bloch equations reduce to:

$$v_z \frac{d\rho}{dz} = -\frac{i}{\hbar}[\mathcal{H}_{tot}, \rho] + \mathcal{L}\rho$$

Taking into account the Doppler frequency shift,

$$\Delta_{pr} = \Delta_{pr}^0 - k_{pr} \cdot v_x$$

$$\Delta_{pu} = \Delta_{pu}^0 + k_{pu} \cdot v_z$$

and by applying the boundary conditions ($(\rho_{11}(t=0) = \rho_{22}(t=0) = 0.5)$), one can derive an effective optical susceptibility for the probe beam, using the following relation:

$$\chi(\Delta_{pu}, \Delta_{pr}) \propto \int_{-\infty}^{\infty} dv_z \int_{-\infty}^{\infty} dv_x \left(\hat{\rho}_{14}(\Delta_{pr}, \Delta_{pu}, v_x, v_z) + \hat{\rho}_{24}(\Delta_{pr}, \Delta_{pu}, v_x, v_z)\right) W(v_x, v_z)$$

where

$$\hat{\rho}_{ij}(\Delta_{pr}, \Delta_{pu}, v_x, v_z) = \int_0^{\infty} \rho_{ij}(\Delta_{pr}, \Delta_{pu}, v_x, v_z, z) dz,$$

and $W(v_x, v_z)$ is the bi-dimensional Maxwellian velocity distribution function,

$$W(v_x, v_z) = \frac{1}{2\pi v_T^2} \exp\left(-\frac{v_x^2 + v_z^2}{2v_T^2}\right)$$

Where $v_x, v_z$ are the atomic velocity along *x* and *z* direction, respectively, and $v_T = \sqrt{k_B T/m}$ is the thermal velocity of the atomic vapor. $k_B$, $T$, and $m$ are the Boltzmann constant, the temperature of the gas and the atomic mass, respectively.

The reflectively for TM polarization of the probe beam is defined as:

$$R_p = |r_p|^2$$

where $r_p$ is the reflection coefficient which can be expressed using Fresnel equations as :

$$r_p = \frac{n_1 \cos\theta_i - \left(\frac{n_1}{n_2}\right)^2 \sqrt{n_2^2 - n_1^2 \sin^2\theta_i}}{n_1 \cos\theta_i + \left(\frac{n_1}{n_2}\right)^2 \sqrt{n_2^2 - n_1^2 \sin^2\theta_i}}$$

Where $n_2$ is the refractive index of Rubidium vapor,

$$n_2(\Delta_{pu}, \Delta_{pr}) = \sqrt{1 + \chi(\Delta_{pu}, \Delta_{pr})}$$

Throughout this article, the reflectivity simulations were carried out using the above equations.

## Experimental Results:

A schematic drawing of the experimental setup is shown in Figure 2(a), including a picture of our Rb-vapor cell made of a Pyrex glass tube terminated one end by a right angle BK-7 prism. The refractive index of the prism is about 1.52 and so the critical angle for total internal reflection is around 42°. A 5 nm thin layer of $MgF_2$ was evaporated on the BK-7 prism before integrating it to the glass tube to avoid chemical reaction between the BK-7 prism and the Rb vapor. The setup was heated using a homemade oven (T~75-80 ºC) with a slight temperature gradient to avoid the condensation of rubidium atoms on the surface of the prism. Further details about the fabrication for the vapor cell can be found in ref. [20].

In the first experiment, the 795 nm pump laser (L1: TOPTICA Photonics, DL 100) is sent in the vapor cell from the top, and at a normal incidence with respect to the surface of the prism. This laser is frequency locked on the $F_g= 2 \rightarrow F_e= 3$ $D_1$ line. The pump beam has a diameter of around 2 mm and an incident power 785 μw. The optical density is around 1, meaning that a significant fraction of the incident power still makes it to the surface of the prism. The 780 nm probe laser (L2: TOPTICA Photonics, DL pro), with an incident power of 24 μw, is sent under total internal reflection conditions (incident angle 43°). Thus, the wave vectors of the pump ($K_{pump}^\perp$) and probe ($K_{probe}^\parallel$) beams are orthogonal to each other. Beam splitter and mirror are used to tap some of the light from the probe laser. This tapping signal is sent through a 7.5 cm long Rb reference cell (Ref. signal). Figure 2(b) shows schematically the interaction of the evanescent field (blue curve) with the atoms at the dielectric-vapor interface.

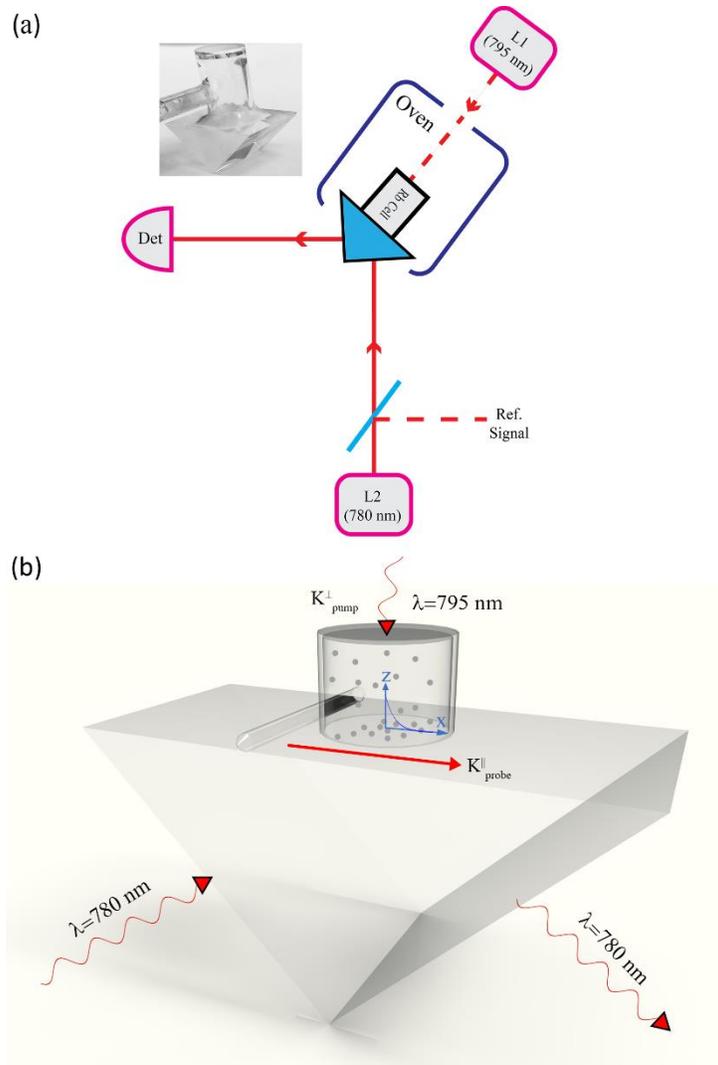

**Figure 2**: (a) Experimental Setup used for optical pumping; Inset shows the photograph of Rb-vapor cell which consists of a Pyrex glass tube and a right angle BK-7 prism (b) Experimental scheme for optical pumping where the wave vectors of the pump and probe beams are perpendicular to each other

Figure 3(a,b) shows the simulated and experimentally measured reflectivity spectra with (red curves) and without (blue curves) optical pumping, respectively. When the pump beam is turned off, we observe four absorption lines corresponding to the transition from both ground states ($F_g$ = 2, 3) to the $F_e$ manifold for both rubidium isotopes, $^{85}$Rb and $^{87}$Rb. The lines at around -4.5 GHz and +2.5 GHz, are due to the existence of $^{87}$Rb. As we focus on the optical pumping in $^{85}$Rb, these lines will be ignored in the forthcoming discussion. Each line has a spectral width of around 600 MHz corresponding to a strong Doppler broadening of the natural $D_2$ line. When the pump beam is turned on, we can clearly see that the absorption from the ground state $F_g$ = 2 vanishes, due to efficient optical pumping of the entire Doppler broadened spectrum.

The basic explanation for this optical pumping is as follows: the resonant beam, propagating along the z axis, pumps all the atoms moving in a plane perpendicular to the beam (i.e. the x-y plane) due to negligible Doppler frequency shift. As for the atoms with significant velocity component along the pump propagation direction, here the nanoscale confinement plays at our favor. Those atoms move away of the evanescent beam with a characteristic time of $(\kappa_{pr} v_T)^{-1} \approx 0.3$ ns, *i.e.* much faster than the natural lifetime ($\approx 20$ ns). As a result, those atoms do not have sufficient time to interact with the probe beam, propagating along the x-axis. Thus, we observe full optical pumping imprinted on the evanescent probe beam, since all the interacting atoms are pumped to the other off-resonant ground state.

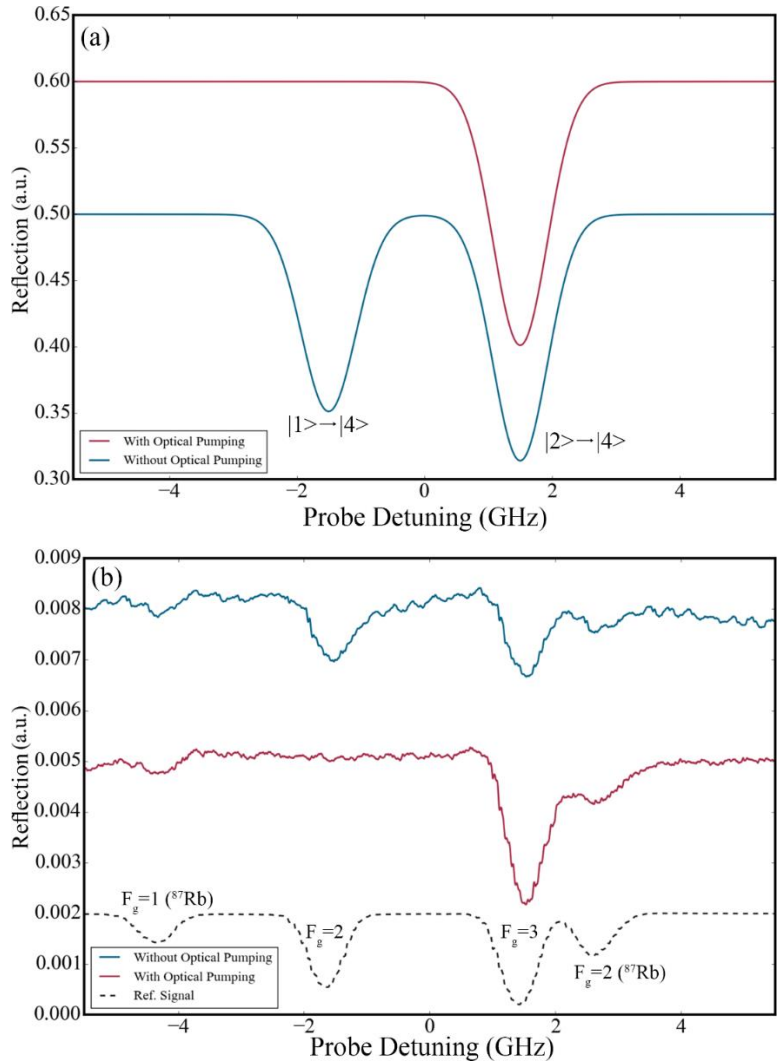

**Figure 3**: (a) Simulated (b) Experimental reflectivity measurements for without and with optical pumping respectively

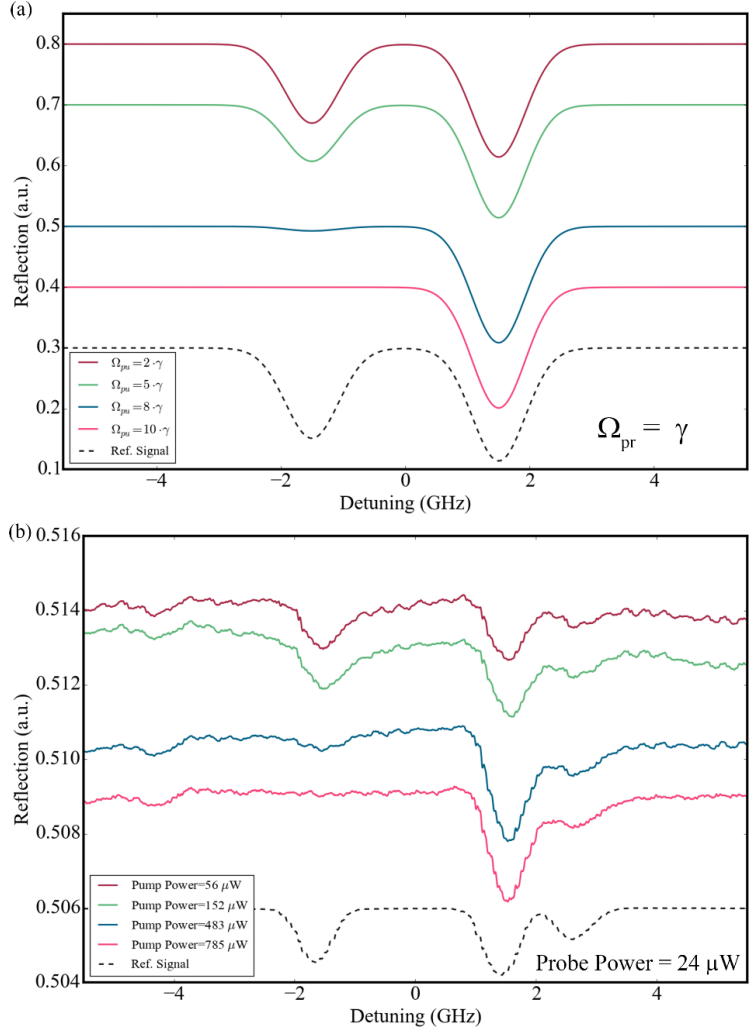

**Figure 4**: (a) Simulated (b) Experimental reflectivity measurements for different intensities of pump beam while the intensity for probe beam was kept constant

To examine the effect of the pump beam intensity on the optical pumping efficiency, we repeated the experiment with different pump beam powers keeping the probe beam power constant (24 µW) as shown in Figure 4a (simulation results) and Figure 4b (experimental results). One can clearly see that the decrease in the pump beam power is followed by a reduction in optical pumping efficiency. This result is explained by the fact that by reducing the pump intensity, fewer atoms are excited from the ground state into the excited state. When reducing the pump beam power to about 100 microwatts and below, no optical pumping can be clearly observed.

To further validate our results, we performed a control experiment, where the pump (795 nm) and the probe (780 nm) beams are co-propagating as shown in Figure 5(a). To detect only the probe beam, two band pass filters were placed before the detector to reject the 795 nm pump beam. The result obtained from co-propagating beam experiment is shown in Figure 5(b). The blue and red lines represent the reflected signal when the pump beam was "off" and "on" respectively. As can be seen, efficient optical pumping cannot be observed. Instead, when the pump beam was on, we could see a small saddle in $F_g = 2$ absorption line which is attributed to velocity selective optical pumping [14].

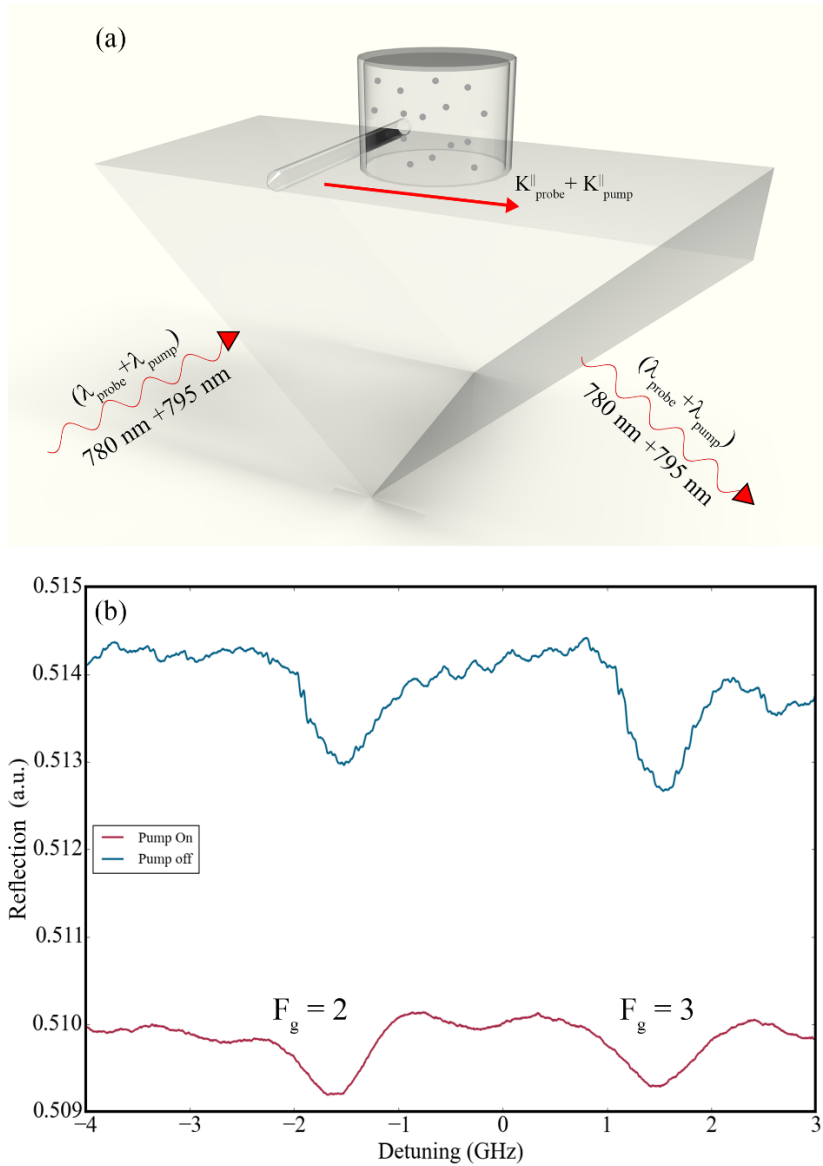

**Figure 5**: (a) Experimental scheme where the pump and probe beam are co-propagating in the prism (b) Experimental reflectivity measurements with pump beam on and off respectively

# Conclusion

In this work, we demonstrated efficient hyperfine optical pumping of an evanescent field at a dielectric-vapor interface using cross beam pump-probe scheme. The experiment was performed with a Rubidium vapor but could be generalized to any alkaline atoms. A miniaturized hand-held device was realized by integrating a right angle prism with a vapor cell. Two pump-probe schemes were presented. In the first configuration, the pump and probe beams were propagating orthogonal to each other, whereas in the other scheme, the pump and probe beam were co-propagated. Highly efficient optical pumping was achieved only with the first configuration. This is explained by the fact that the pump beam could excite almost all the atoms moving perpendicular to the pump beam due to negligible Doppler shift. Additionally, atoms with significant velocity component along the pump beam, which cannot be pumped due to Doppler shift, contribute marginally to absorption as they quickly move away of the submicron confined evanescent probe beam. The results were obtained under different pump beam intensities and a decrease in the contrast of optical pumping was noticed with the decrease in the pump beam intensity. To support our experimental observations, we have calculated the expected reflected signal in our system using numerical simulations based on optical Bloch equations. The simulation results were found in good agreement with the observed experimental results. Finally, we believe that the obtained on-chip highly efficient optical pumping at nanoscale serves as an important step in the quest for realizing miniaturized quantum devices for diverse applications ranging from optical switching to magnetometry and quantum repeaters.